\documentclass[12pt]{article}
\usepackage{epsfig, amssymb}
\newcommand{\be}{\begin{equation}}
\newcommand{\ee}{\end{equation}}
\newcommand{\bea}{\begin{eqnarray}}
\newcommand{\eea}{\end{eqnarray}}
\newcommand{\bA}{\begin{array}}
\newcommand{\eA}{\end{array}}
\newcommand{\bc}{\begin{center}}
\newcommand{\ec}{\end{center}}
\newcommand{\al}{\alpha}

\newcommand{\ra}{\rightarrow}
\newcommand{\del}{\partial}

\newcommand{\ie}{{\it i.e.}}
\newcommand{\eg}{{\it e.g.}}

\newcommand{\Rea}{\mathop{\rm Re}}

\newcommand{\Nt}{${\cal N}{=}2$}

\def\BC{{\mathbb C}}
\def\BP{{\mathbb P}}
\def\BR{{\mathbb R}}
\def\BZ{{\mathbb Z}}
\def\BQ{{\mathbb Q}}

\def\BN{\mbox{\boldmath$N$}}

\newcommand{\ov}{\over}

\begin{document}

\begin{titlepage}
%\vspace{10mm}

\bc

\hfill  {TIFR/TH/05-42} \\
\hfill  {NSF-KITP-05-83} \\
\hfill  {\tt hep-th/0510104} \\
         [22mm]
%X\vfill

{\Huge Closed string tachyons, flips and conifolds}
\vspace{10mm}

{\large K.~Narayan} \\
\vspace{3mm}
{\small \it Department of Theoretical Physics, \\}
{\small \it Tata Institute of Fundamental Research, \\}
{\small \it Homi Bhabha Road, Colaba, Mumbai - 400005, India.\\}
{\small Email: \ narayan@theory.tifr.res.in}\\

\ec
\medskip
\vspace{40mm}

\begin{abstract}
Following the analysis of tachyons and orbifold flips described in
hep-th/0412337, we study nonsupersymmetric analogs of the
supersymmetric conifold singularity and show using their toric
geometry description that they are nonsupersymmetric orbifolds of the
latter. Using linear sigma models, we see that these are unstable to
localized closed string tachyon condensation and exhibit flip
transitions between their two small resolutions (involving 2-cycles),
in the process mediating mild dynamical topology change. Our analysis
shows that the structure of these nonsupersymmetric conifolds as
quotients of the supersymmetric conifold obstructs the 3-cycle
deformation of such singularities, suggesting that these
nonsupersymmetric conifolds decay by evolving towards their stable
small resolutions.
\end{abstract}

\end{titlepage}

\newpage 
{\small
\begin{tableofcontents}
\end{tableofcontents}
}

\vspace{2mm}

\section{Introduction}

The study of closed string tachyon condensation in unstable geometries
in string theory, beginning with \cite{aps}, followed by
\cite{vafa0111}, \cite{hkmm}, and others, has been quite rich (see
\eg\ the reviews \cite{emilrev} \cite{minwalla0405}), from the point
of view of understanding the vacuum structure of string theory, the
role of time, as well as phenomena pertaining to quantum geometry.
While direct (second order) time evolution is often hard to study,
considerable qualitative insight is gained by studying (first order)
renormalization group flows in appropriate auxiliary 2-dimensional
(worldsheet) theories describing the unstable system: these are often
tractable when worldsheet supersymmetry is unbroken, spacetime
supersymmetry being broken, and are then closely intertwined with the
algebro-geometric structure of the singularities and their
resolutions. Closed string tachyons localized at singularities,
somewhat analogous to open string tachyons localized on
brane-antibrane defects in a fixed ambient spacetime, are simpler than
delocalized instabilities since initially they affect only the
vicinity of the singularity: in fact, their condensation often turns
out to resolve the singularity.

Closed string tachyons localized at $\BC^3/\BZ_N$ nonsupersymmetric
orbifold singularities were studied in \cite{drmknmrp} (see also
\cite{sarkar0407}), with a more detailed gauged linear sigma model
(GLSM) analysis of dynamical topology change via flip transitions
given in \cite{drmkn}. Codimension three singularities are somewhat
complicated due to the existence of terminal singularities and the
absence of canonical resolutions. The analysis of possible endpoints
in \cite{drmknmrp} involved studying the decay channel corresponding
to the condensation of the most relevant tachyon and sequentially
iterating this procedure for each of the residual geometries, which
are themselves typically unstable. The generic endpoints of decay for 
Type II string propagation on such singularities are smooth, \ie\ 
spaces with at worst supersymmetric singularities that can be resolved 
by moduli (marginal operators); however, Type 0 theories do in fact 
generically exhibit a terminal singularity $\BC^3/\BZ_2 (1,1,1)$ in 
their spectrum of decay endpoints. Generically there are multiple
distinct tachyons. This is particularly important for codimension three
singularities due to the absence of canonical resolutions: the various
possible distinct resolutions are related by flips and flops, mediated
by relevant and marginal operators, respectively, in the worldsheet
theory. The competition between tachyons of distinct R-charges (\ie\
masses in spacetime) gives rise to flips \cite{drmkn}, described in
part in \cite{drmknmrp}, relating the distinct partial resolutions of
the original singularity, thought of as distinct attractor basins for
the worldsheet RG flow. A flip transition can be thought of as a
blowdown of a cycle accompanied by a blowup of a topologically
distinct cycle: this involves a mild change in the topology of the
ambient (embedding) compact space containing such local singularities,
with changes in the intersection numbers of the various cycles of 
the geometry, the two cycles being topologically distinct (although 
the analogs of the Hodge numbers, \ie\ the numbers of $p$-cycles,
themselves do not change, as for supersymmetric flops in Calabi-Yau
spaces \cite{psabrgdrm} \cite{wittenphases}).
Orbifold flips \cite{drmkn} can always be consistently embedded in a
2-tachyon sub-GLSM of the full, say, $n$-tachyon GLSM, reflecting the
fact that in the corresponding toric fan, they occur in subcones
containing a reversal of the sequence of subdivisions pertaining to
only one wall in the fan: we will refer to this subcone containing the
nontrivial flip dynamics as a $flip$ $region$. Physically a flip
occurs when a more dominant tachyon condenses during the condensation 
of some tachyon, and the effective dynamics of the
relative condensation process can be described by a GLSM with gauge
group $U(1)$, as we will review later. In particular the dynamics of
the GLSM RG flow always drives a flip transition in the direction of
the partial resolution leading to a less singular residual geometry,
which can be thought of as a more stable endpoint: this enables a
classification of the phases of GLSMs corresponding to these unstable
orbifolds into ``stable'' and ``unstable'' basins of attraction,
noting the directionality of the RG trajectories involving potential
flip transitions, which always flow towards the more stable phases.\\
We mention here in passing that D-branes and Coulomb branches in
$\BC^2/\BZ_N$ singularities have been studied most recently in
\cite{moore0507} following \cite{martinecmoore, moore0403}, while
\cite{immrp0501, immrp0507} study the emergence of Coulomb branches in
codimension three. \cite{adams0502} describes topology-changing
transitions mediated by tachyons in string compactifications on 
Riemann surfaces. See also \eg\ \cite{mcgreevysilverstein0506, 
zwiebach0506076, zwiebach0506077, horowitz0506, ross0509, kleb0509,
kutasov0509, silver0510, bergman0510}, for other interesting recent
work pertaining to closed string tachyons.

As mentioned above (and will be reviewed in more detail in Sec.~2), 
an orbifold flip occurs within a flip region representing a 2-tachyon
sub-GLSM and the condensation of the tachyons ``relative'' to each
other can be described by an effective GLSM with gauge group $U(1)$:
in this case, the effective dynamics of the flip region is simply a
part of the full decay structure of the orbifold, the corresponding
subcone being embedded in the full toric fan of the orbifold. In this
work, we study these flip regions as singularities in themselves,
devoid of the orbifold embedding, and analyze their dynamics. From
their toric data, we show that they are unstable conifold-like
singularities, obtained as nonsupersymmetric orbifolds of the
supersymmetric conifold. They are labelled by a charge matrix
\be
Q=(\bA{cccc} n_1 & n_2 & -n_3 & -n_4 \eA)\ , \qquad\ \ 
\Delta n\equiv\sum Q_i\neq 0\ ,
\ee
for integers $n_i>0$, which characterizes their toric data and
therefore their geometry (in this notation, the supersymmetric
conifold corresponds to $Q=(\bA{cccc} 1 & 1 & -1 & -1 \eA)$). As
orbifolds of the supersymmetric conifold, they can be described by a
hypersurface equation $z_1 z_4 - z_2 z_3 = 0$, with the new feature
that the $z_i$ are coordinates not in $\BC^4$ but in $\BC^4/\Gamma$
where $\Gamma$ is a discrete group depending on the $n_i$. Throughout
our discussion, we view such a geometry as the vicinity of a local
3-complex dimensional singularity embedded in some compact space
(possibly some appropriate nonsupersymmetric orbifold of a Calabi-Yau
that develops a supersymmetric conifold singularity) and focus on the
local dynamics of the singularity, the full spacetime in this
effective noncompact limit being of the form $\BR^{3,1}\times {\cal
C}^{(flip)}$.  With a view to understanding the dynamics in our
nonsupersymmetric case here, somewhat analogous to \cite{strominger}
\cite{greemorrstro} (see also the review \cite{greeneCY}) for the
supersymmetric conifold, we study the geometry in the vicinity of the
singularity in Sec.~3, in part drawing analogies with the
corresponding analysis \cite{candelasconif} of the supersymmetric
conifold. We recall that the supersymmetric conifold admits two
topologically distinct small resolutions replacing the singularity by
2-cycles (K\"ahler deformations related by a flop) and a complex
structure deformation replacing the singularity by a 3-cycle. In the
case at hand, our analysis shows that these 3-cycle deformations are 
obstructed for such flip conifold singularities due to their structure 
as quotients of the supersymmetric conifold\footnote{Note 
that this does not however preclude abstract deformations.}, suggesting 
that the decay structure of these singularities is always via
their small resolutions, involving 2-cycles. Unlike a flop, the fact
that $\sum Q_i\neq 0$ gives rise to an asymmetry in the resolutions,
so that one of the resolutions has a spontaneous tendency to evolve
towards the other, more stable, one: thus there is an inherent
directionality here, leading to dynamical topology change with a
blowdown of a 2-cycle followed by a spontaneous blowup of the other
(topologically distinct) 2-cycle. We study the dynamics of the small
resolutions of these singularities using linear sigma models, gaining 
insight from the RG dynamics of the Fayet-Iliopoulos parameter. As in
orbifold decay where divisors blow up and expand in time, potentially
containing unstable residual singularities on their loci, the
2-cycles in the small resolution decay channels of these flip
conifolds potentially contain unstable residual orbifold singularities
which then themselves decay. Using the Type II GSO projection for
these residual orbifold singularities, we find a nontrivial
constraint\ $\Delta n=even$ \ for the $\BR^{3,1}\times {\cal
C}^{(flip)}$ spacetime background to admit a Type II GSO projection
with no bulk tachyons: this is consistent with the endpoint orbifold
singularities also admitting Type II GSO projections. Finally in
Sec.~4 we conclude with a brief discussion on the evolution of these
geometries.

\section{Flip transitions in $\BC^3/\BZ_N$ orbifolds}

The closed string tachyon dynamics of nonsupersymmetric $\BC^3/\BZ_N$
orbifolds has been studied in some detail in \cite{drmknmrp}
\cite{drmkn}\footnote{See also \cite{sarkar0407} \cite{immrp0501}.}.
The absence of canonical (minimal) resolutions gives rise to flips.

\begin{figure}
\bc
\epsfig{file=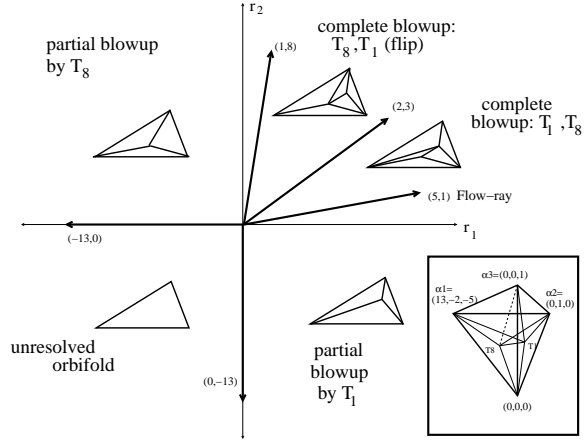, width=7.6cm}
\caption{The phase structure of $\BC^3/\BZ_{13} (1,2,5)$ for two 
tachyons. The inset shows the toric fan and its subdivisions.}
\label{figphases}
\ec
\end{figure}

To illustrate these phenomena, we review a concrete example 
$\BC^3/\BZ_{13} (1,2,5)$ and the flip dynamics here, studied in detail 
in \cite{drmkn}.
This can be described by a GLSM with gauge group $U(1)\times U(1)$ 
and two of five chiral superfields representing tachyons $T_1$ and 
$T_8$ with masses $m_1^2={2\over\al'}({8\over 13}-1)$ and
$m_8^2={2\over\al'}({11\over 13}-1)$ respectively, their R-charges 
given by $R_1\equiv ({1\over 13},{2\over 13},{5\over 13})
={8\over 13}$ and $R_8\equiv ({8\over 13},{3\over 13},{1\over 13})
={12\over 13}$. The action of the GLSM is (we use the conventions of 
\cite{wittenphases, morrisonplesserInstantons})
\be
S = \int d^2 z\ \biggl[ d^4 \theta\ \biggl( {\bar \Psi_i} e^{2Q_i^a 
V_a} \Psi_i - {1\over 4e_a^2} {\bar \Sigma_a} \Sigma_a \biggr) + 
\Rea\biggl( i t_a\int d^2 {\tilde \theta}\ \Sigma_a  \biggr) \biggr]\ ,
\ee
summation over $a=1,2$, being implied. 
The\ $t_a = ir_a + {\theta_a\over 2\pi}$ \ are Fayet-Iliopoulos 
parameters and $\theta$-angles for both the gauge fields, $e_a$ being 
the gauge couplings. The twisted chiral superfields $\Sigma_a$ (whose 
bosonic components include complex scalars $\sigma_a$) represent 
field-strengths for the gauge fields. The action of the $U(1)^2$ on 
the chiral superfields $\Psi\equiv \phi_1, \phi_2, \phi_3, T_1, T_8$, 
comprising the three coordinate superfields and the two tachyons, is 
given by $\Psi_i \ra e^{i Q_i^a\lambda}\ \Psi_i$, where the charge 
matrix is
\be\label{Qia2gen}
Q_i^a = \left( \bA{ccccc} 1 & 2 & 5 & -13 & 0  \\ 8 & 3 & 1 
& 0 & -13  \\ \eA \right), \qquad i=1,\ldots,5, \ \ a=1,2\ .
\ee
Such a charge matrix only specifies the $U(1)\times U(1)$ action up to 
a finite group, due to the possibility of a $\BQ$-linear combination 
of the rows of the matrix also having integral charges.
There are two independent FI parameters (for the two $U(1)$'s) whose 
variations control the vacuum structure of the theory. The space of 
classical ground states of this theory can be found from the bosonic 
potential 
\be\label{bospot}
U = \sum_a {(D_a)^2\over 2e_a^2} + 2\sum_{a,b=1}^2 {\bar 
\sigma}_a \sigma_b \sum_i Q_i^a Q_i^b |\Psi_i|^2\ .
\ee
Then $U=0$ requires $D_a=0$: solving these for $r_a\neq 0$ gives 
expectation values for the $\Psi_i$, which Higgs the gauge group down 
to some discrete subgroup and lead to mass terms for the $\sigma_a$ 
whose expectation values thus vanish. The classical vacua of the 
theory are then given in terms of solutions to the D-term equations
\be
{-D_a\over e^2} = \sum_i Q_i^a |\Psi_i|^2 - r_a = 0\ , \qquad a=1,2\ .
\ee
These give collections of coordinate charts that characterize in
general distinct toric varieties. In other words, this 2-parameter
system admits several ``phases'' (convex hulls in $r$-space) depending 
on the values of $r_1, r_2$, shown in Figure~\ref{figphases}. 
In this case, there are five phases, corresponding to the unresolved
orbifold, the partial resolutions by condensation of $T_1$ or $T_8$
alone, and the complete resolutions by condensation of both $T_1$ and
$T_8$. Note that the order in which the tachyons condense changes the
resulting endpoint: condensation of $T_8$ followed by $T_1$ gives rise
to an endpoint geometry that is related by a flip transition
(blowdown+blowup) to that obtained by condensation of the sequence
$T_1, T_8$. Each phase is an endpoint since if left unperturbed, the
corresponding resolution of the singularity can continue indefinitely
(within this noncompact approximation): in this sense, each phase 
(more accurately, an appropriate limit inside the phase) is a fixed 
point of the GLSM RG flow. However some of these phases are unstable 
while others are stable, in the sense that small fluctuations
(condensation of other tachyons) will cause the system to run away
from the unstable phases towards the stable ones. This can be gleaned
from the 1-loop renormalization of the FI parameters
\be
r_a = \bigg({\sum_iQ_i^a\over 2\pi}\bigg) \log {\mu\over \Lambda}\ ,
\ee
where $\mu$ is the RG scale and $\Lambda$ is an ultraviolet cutoff 
scale where the $r_a$ are defined to vanish. We restrict attention 
to the large $r_a$ regions, thus ignoring worldsheet instanton 
corrections, which is sufficient for our purposes of understanding 
the phase structure. A generic linear combination of the gauge 
fields couples to a linear combination of the FI parameters and, in 
this example, has the 1-loop running
\be\label{flowray51}
c_1 r_1 + c_2 r_2 = -\biggl({5c_1+c_2\over 2\pi}\biggr) 
\log {\mu \over \Lambda}\ .
\ee
This parameter is marginal if \ $5 c_1 + c_2 = 0$ : this describes 
a line perpendicular to the ray $(5,1)$ in $r$-space (\ie\ the
secondary fan). Thus we can choose the basis for the $U(1)\times U(1)$
so that one linear combination couples to a (relevant) FI parameter
that has nontrivial renormalization along the flow, while the other FI
parameter is marginal along the flow. From (\ref{flowray51}), we see
that this single relevant direction (perpendicular to the marginal
one), lies along the ray $(5,1)$: this is the flow-ray. These
equations indicate that at low energies $\mu\ll\Lambda$, the 1-loop 
RG flows drive the system towards the large $r$ regions in $r$-space,
\ie, $r_1, r_2\gg 0$, that are adjacent to the flow-ray $(5,1)$: 
these phases, \ie\ the convex hulls $\{(5,1),(0,-13) \}$\ and 
$\{(2,3),(5,1)\}$, are thus the stable attractors for the GLSM RG flow. 
For generic combinatorics, the flow-ray lies in the interior of some
convex hull in the secondary fan, in which case there is precisely 
one stable phase. However if the flow-ray is the boundary of two 
adjacent convex hulls, as it is here, there will be multiple stable 
phases related by marginal deformations or moduli in the infrared. 
From Figure~\ref{figphases}, we see that the two stable phases arising 
as the endpoints of flowlines correspond to the partial resolution by 
the tachyon $T_1$ alone and the complete resolution by the tachyon 
$T_1$ followed by $T_8$. This dovetails nicely with the toric analysis 
in \cite{drmknmrp}: after condensation of $T_1$ (the most relevant 
tachyon here), the subsequent tachyon $T_8$ becomes massless 
so that the blowup it corresponds to is now a modulus or flat 
direction, rather than a residual instability. This marginality of 
the residual (erstwhile tachyon) $T_8$ reflects the marginal $U(1)$ 
direction perpendicular to the flow-ray $(5,1)$ in the GLSM. Thus 
the GLSM RG flow drives the system along the flow-ray, maintaining 
a flat direction perpendicular to it.

In the process of ending up in one of these stable phases, flowlines
may cross one or more of the the (semi-infinite) phase boundaries
emanating from $(0,0)$ and passing through $(-13,0), (1,8), (2,3)$ and
$(0,-13)$. In particular in crossing $(2,3)$, a flip occurs, involving
a reversal of the sequence of partial resolutions of the orbifold by
the two tachyons, from $T_8$ followed by $T_1$, to $T_1$ and then
$T_8$.  The physics of this phase boundary occurs in the region of the
2D moduli space where the nontrivial dynamics does not involve the
field $\phi_2$, enabling us to study the effective dynamics of the
flip transition captured by this subsystem, the flip region here being
defined by the subcone $T_1, T_8, \phi_1, \phi_3$ (see 
Figure~\ref{figphases}). This is described by an effective $U(1)$ GLSM 
with the effective D-term
\be
{-D^{eff}\over 13e^2} = -{1\over 13}(3D_1-2D_2) = 
|\phi_3|^2 + 2|T_8|^2 - |\phi_1|^2 - 3|T_1|^2 - r^{eff} = 0\ ,
\ee
the effective charge matrix being 
$Q^{eff}=(\bA{cccc} 1 & 2 & -1 & -3 \eA)$ . This effective 
Fayet-Iliopoulos parameter $r^{eff}={3r_1-2r_2\over 13}$ has a 1-loop 
renormalization given by 
\be
r^{eff} = (\sum_i Q_i^{eff})\ \log {\mu\over \Lambda} 
= (-1)\ \log {\mu\over \Lambda}\ ,
\ee
showing the inherent directionality of the flip in spacetime.
Thus $r^{eff}$ flows under the GLSM renormalization group from the 
$r^{eff}\ll 0$ phase (partial resolution by a 2-cycle with coordinate 
charts $(\phi_3,T_8,\phi_1), \ (\phi_3,T_8,T_1)$) to the
$r^{eff}\gg 0$ phase (partial resolution by a different 2-cycle with 
charts $(\phi_3,\phi_1,T_1), \ (T_8,\phi_1,T_1)$) which has distinct 
topology. 

This is a fairly generic story for flip dynamics in unstable
orbifolds, described in generality in \cite{drmkn}. We recall
\cite{drmknmrp} \cite{drmkn} that the cumulative degree of the
residual singularities after condensation of a tachyon $T_j$ with
R-charge $R_j$, \ie\ subdivision by the corresponding lattice point 
in the toric cone, is given by the total $\BN$ lattice volume 
$V(T_j)=NR_j$ of the residual subcones: thus a more relevant tachyon
gives rise to a smaller $\BN$ lattice volume, \ie\ a less singular
residual endpoint. The difference in the subcone volumes in the two 
partial resolutions (see Figure~\ref{figphases})
\be\label{voldiffOrbs}
\Delta V = V(0;\al_1,T_1,\al_3) + V(0;\al_3,T_1,T_8) - V(0;\al_1,T_1,T_8) 
- V(0;\al_1,T_8,\al_3)\ ,
\ee
can be thought of as a quantitative measure of the difference in the 
degree of singularity of the two resolutions. Furthermore the 
coefficient of the logarithm in $r^{eff}$ turns out to be
\be\label{rflowOrbs}
r^{eff} = (\Delta V)\cdot \log {\mu \over \Lambda}\ .
\ee
Thus the RG flow for this effective FI parameter proceeds precisely in
the direction of decreasing $\BN$ lattice volume, \ie, in the
direction along which $\Delta V < 0$. In other words, the GLSM RG flow
dynamics drives the flip transition in the direction of the less
singular residual geometry, which in this example is the stable phase
corresponding to the blowup sequence $T_1, T_8$.

The 1-loop renormalization of the FI parameters can be expressed 
\cite{wittenphases, wittenIAS, morrisonplesserInstantons} in terms of 
a perturbatively quantum-corrected twisted chiral superpotential for 
the $\Sigma_a$ 
\be\label{twistedW}
{\tilde W}(\Sigma) = {1\over 2\sqrt{2}} \sum_{a=1}^n \Sigma_a \bigg( 
i{\hat \tau}_a - {1\over 2\pi} \sum_{i=1}^{d+n} Q_i^a \log (\sqrt{2} 
\sum_{b=1}^n Q_i^b \Sigma_b/\Lambda) \bigg)\ 
\ee
for a general $n$-tachyon system, obtained by considering the 
large-$\sigma$ region in field space and integrating out those scalars 
$\Psi_i$ that are massive here (and their expectation values vanish 
energetically). This leads to the modified potential 
\be\label{twistedWpot}
U(\sigma) = {e^2\over 2} \sum_{a=1}^n \bigg| i{\hat \tau}_a - 
{\sum_{i=1}^{d+n} Q_i^a \over 2\pi} (\log (\sqrt{2} \sum_{b=1}^n Q_i^b 
\sigma_b/\Lambda) + 1) \bigg|^2\ .
\ee
The singularities predicted classically at the locations of the phase 
boundaries arise from the existence of low-energy states at large 
$\sigma$. Now from above, we see that along the single relevant 
direction where $\sum_i Q_i^1\neq 0$, the potential energy has a 
$|\log (\sigma_1)|^2$ growth so that the field space accessible to very 
low-lying states is effectively compact and there is no singularity 
along the single relevant direction given by the flow-ray \cite{drmkn}: 
in other words, the RG flow is smooth along the tachyonic directions 
for all values of $\tau_1$, and the phase boundaries are simply labels 
for the boundaries of the adjacent phases.

\section{Tachyons and flip conifolds}

In the previous section, the flip regions in question were embedded
within nonsupersymmetric $\BC^3/\BZ_N$ orbifold singularities. In what
follows, we will study the singularity structure of flip regions
treating them as geometric objects in their own right, described by
their toric data. In particular, from the toric fan of a given flip
region, we can glean the algebraic structure of the corresponding
singularity. We will see that these flip regions are to be thought of
as nonsupersymmetric analogs of the supersymmetric conifold
singularity. 

Let us first recall some key features of the supersymmetric conifold
(the description below of the toric data, the corresponding $U(1)$ 
action and algebraic singularity structure for the supersymmetric 
conifold can be found in \eg\ \cite{morrisonplesserHorizons}):
this can be represented by a $\BC^*$ action given by the charge matrix
\be\label{susy1111}
Q = \left( \bA{cccc} 1 & 1 & -1 & -1 \eA \right)\ ,
\ee
with action\ $\Psi_i\ra \lambda^{Q_i} \Psi_i, \ \lambda\in\BC^*$, on 
the coordinates $\Psi_i\equiv a, b, c, d$, describing the singularity.
The supersymmetric conifold is specified by toric data (see 
Figure~\ref{figflip}) given by four (minimal) lattice points $e_i$ such 
that $e_1+e_2-e_3-e_4=0$, the coefficients being fixed by the charges 
$Q_i$. A basis of monomials invariant under the $\BC^*$ action is\ 
$z_1=ac,\ z_2=ad,\ z_3=bc,\ z_4=bd$, satisfying
\be\label{11-1-1}
z_1 z_4 - z_2 z_3 = 0\ ,
\ee
which describes the supersymmetric conifold as a hypersurface\footnote{
This can be recast as $\sum_{i=1}^4 w_i^2 = 0$, by appropriate 
redefinitions of the coordinates $z_i$.} embedded in $\BC^4$. 
%Note that each of the two terms on the left hand side in (\ref{11-1-1}) 
%has total $\BC^*$ charge zero, \ie\ the $U(1)$ action on each term is 
%trivial. 
Now recall that 
the supersymmetric conifold singularity above can be smoothed out in
two distinct ways:\ $(i)$ via either of the two small resolutions of
the singularity obtained by blowing up two-spheres ($\BP^1$s) at the
singularity -- these are K\"ahler deformations, related by a flop
transition; \ $(ii)$ via a complex structure deformation of the
hypersurface equation as\ $z_1 z_4 - z_2 z_3 = \epsilon$ , obtained by
blowing up a three-sphere. 

In what follows, we study along the above lines the geometry of flip
regions.

\subsection{An example:\ $(1\ \ 2\ -1\ -3)$}

For instance, let us revisit the flip region in $\BC^3/\BZ_{13}\ (1,2,5)$: 
a basis for the $\BN$ lattice containing the tetrahedral flip region 
consists of the vertices\ 
$e_3 \equiv \phi_3=(0,0,1), \ e_1\equiv T_1=(1,0,0), \ 
e_2\equiv T_8=(0,1,0)$, relabelling the tachyonic lattice point 
$T_8=(8,-1,-3)$ for convenience (see Figure~\ref{figphases}). Then 
the coordinate vertex $\phi_1=(13,-2,-5)\equiv e_4$ in terms of the 
$e_1,e_2,e_3$ basis satisfies the relation\ $e_4+3e_1-2e_2-e_3=0$, which 
is suggestively reminiscent of the equation describing the supersymmetric 
conifold as a hypersurface in $\BC^4$. Let us now strip off the orbifold 
embedding and study this flip region as a geometric object in itself 
(see Figure~\ref{figflip}): with this in mind, let us realize the 
algebraic structure of the flip from its toric data. 
The $\BC^*$ action here on $a,b,c,d$, is specified by the charge matrix
\be
Q = \left( \bA{cccc} 1 & 2 & -1 & -3 \eA \right)\ .
\ee
Then the redefined coordinates\ $a,\ b^{1\ov 2},\ c,\ d^{1\ov 3}$ \ 
have a new $\BC^*$ action specified by the matrix\ 
$Q = ( \bA{cccc} 1 & 1 & -1 & -1 \eA )$. Defining the monomials\ 
$z_1=ac ,\ z_2=ad^{1\ov 3} ,\ z_3=b^{1\ov 2}c ,\ z_4=b^{1\ov 2}d^{1\ov 3}$, 
invariant under the new $\BC^*$ action, it is straightforward to see that 
the $z_i$ satisfy the relation\ $z_1 z_4 - z_2 z_3 = 0$, \ie\ the 
conifold equation (\ref{11-1-1}).

However note that now there is a residual discrete $\BZ_6$ acting 
nontrivially on the $z_i$, stemming from the well-defined geometric 
rotation by $e^{2\pi i}$ on the coordinates $a, b, c, d$. 
The action of the $\BZ_6$ is
\be\label{12-1-3}
( \bA{cccc} z_1 & z_2 & z_3 & z_4 \eA )\ \ra\ 
( \bA{cccc} z_1 & \omega^{1/3}z_2 & \omega^{1/2}z_3 & \omega^{5/6}z_4 \eA )\ ,
\qquad \qquad \omega=e^{2\pi i}\ ,
\ee
which can be rewritten in terms of $\omega'=e^{2\pi i/6}$. In other 
words, the $z_i$ as coordinates are well-defined only up to these
discrete identifications: we will elaborate on this description in the 
next subsection. Thus the geometry of this flip conifold is somewhat 
different from the supersymmetric conifold, which has no such discrete 
symmetry. In this case, the flip conifold is a hypersurface in 
$\BC^4/\BZ_6$ with the $\BZ_6$ action (\ref{12-1-3}). This is a 
generic feature of such tachyonic conifolds exhibiting 
nonsupersymmetric flip regions, as we will study in greater
detail below.

From (\ref{12-1-3}), we can see that each of the two terms in the
hypersurface equation (\ref{11-1-1}) for the flip conifold above has a
nontrivial action of the $\BZ_6$, given by the phase $e^{2\pi i(5/6)}$. 
Thus there is an obstruction to the existence of a 3-cycle ($S^3$)
deformation of the hypersurface equation of the form\
$z_1 z_4 - z_2 z_3 = \epsilon\neq 0$ , since the deformation parameter 
$\epsilon$ can be chosen a real number (by coordinate changes) trivial
under $\BZ_6$ and does not respect the $\BZ_6$ symmetry of the
original singularity. Roughly speaking, such a deformed geometry lies
``outside'' the symmetry-preserving phases of the geometry that are
connected to the singular point itself. 

In what follows, we will study the geometry of arbitrary flip 
conifolds using their toric data along the lines above.

\subsection{The geometry of the\ $(n_1\ \ n_2\ -n_3\ -n_4)$ \ flip 
conifold region}

\begin{figure}
\bc
\epsfig{file=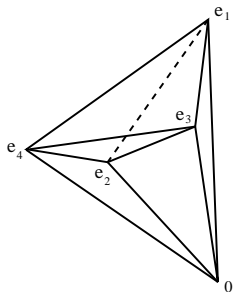, width=3cm}
\caption{The toric fan for a flip region alongwith the two small 
resolutions.}
\label{figflip}
\ec
\end{figure}

Consider a charge matrix 
\be\label{Qgen}
Q = \left( \bA{cccc} n_1 & n_2 & -n_3 & -n_4 \eA \right)
\ee
and a $\BC^*$ action on the complex coordinates 
$\Psi_i\equiv a, b, c, d$, with this charge matrix as 
$\Psi_i\ra \lambda^{Q_i}\Psi_i, \ \lambda\in\BC^*$. 
Since we have defined the minus signs to be in specific places in $Q$, 
the $\{n_i\}$ must be treated as an $ordered$ set. For an unstable 
(tachyonic) geometry, we have\ $\sum_i Q_i\neq 0$. 
The flip region corresponding to this $Q$ can be described, as in 
Figure~\ref{figflip}, by a toric cone defined by the lattice vectors 
$e_i$ satisfying the relation 
\be\label{fan}
\sum Q_i e_i = n_1 e_1 + n_2 e_2 - n_3 e_3 - n_4 e_4 = 0 
\ee
in a 3-dimensional $\BN$ lattice. Then the redefined coordinates 
$a^{1\ov n_1},\ b^{1\ov n_2},\ c^{1\ov n_3},\ d^{1\ov n_4}$ \ have a new 
$U(1)$ action given by\ ${\tilde Q} = ( \bA{cccc} 1 & 1 & -1 & -1 \eA )$ .
Defining the invariant monomials
\be\label{zin1n2n3n4}
z_1=a^{1\ov n_1}c^{1\ov n_3} ,\ \ \ z_2=a^{1\ov n_1}d^{1\ov n_4} ,\ \ \ 
z_3=b^{1\ov n_2}c^{1\ov n_3} ,\ \ \ z_4=b^{1\ov n_2}d^{1\ov n_4}\ ,
\ee
we see that the $z_i$ satisfy
\be\label{flipconif}
z_1 z_4 - z_2 z_3 = 0\ ,
\ee
\ie\ the conifold equation (\ref{11-1-1}). Each of the two terms on the 
left hand side of this conifold equation now has acting on it a discrete 
phase, 
\be\label{phasenu}
e^{2\pi i \biggl({1\ov n_1}+{1\ov n_2}+{1\ov n_3}+{1\ov n_4}\biggr)} 
\equiv e^{2\pi i \nu}\ ,
\ee
stemming from the simultaneous geometric rotation by $e^{2\pi i}$ on 
the\ $\Psi_i\equiv a, b, c, d \in \BC^4$, recast in terms of the $z_i$. 
The phases $e^{2\pi i/n_k}$ induced on the $z_i$ by the independent 
rotations on the underlying variables $a, b, c, d$, determine the 
geometry of deformations of the conifold singularity itself. These 
induce a quotient structure on the flip region with a discrete group 
$\Gamma$ (analogous to the $\BZ_6$ earlier): the coordinates $z_i$ 
have the identifications 
\bea\label{Zn1n2n3n4}
( \bA{cccc} z_1 & z_2 & z_3 & z_4 \eA )\ &\longrightarrow^{a}&
( \bA{cccc} e^{2\pi i/n_1}z_1 & e^{2\pi i/n_1}z_2 & z_3 & z_4 \eA )\ ,
\nonumber\\
&\longrightarrow^{b}&
( \bA{cccc} z_1 & z_2 & e^{2\pi i/n_2}z_3 & e^{2\pi i/n_2}z_4 \eA )\ ,
\nonumber\\
&\longrightarrow^{c}&
( \bA{cccc} e^{2\pi i/n_3}z_1 & z_2 & e^{2\pi i/n_3}z_3 & z_4 \eA )\ ,
\\
&\longrightarrow^{d}&
( \bA{cccc} z_1 & e^{2\pi i/n_4}z_2 & z_3 & e^{2\pi i/n_4}z_4 \eA )\ ,
\nonumber
\eea
under the independent rotations on each of $a,b,c,d$. If the $n_i$ 
are coprime, then this orbifold action can be described by a cyclic 
group $\BZ_N=\prod_i \BZ_{n_i}$, where $N={\rm lcm}(n_1,n_2,n_3,n_4)$ 
is the least common multiple of the $n_i$. 
Thus in general the flip conifold ${\cal C}^{(flip)}$ described by\ 
$Q=(\bA{cccc} n_1 & n_2 & -n_3 & -n_4 \eA)$ is the quotient\footnote{See 
\eg\ \cite{hubsch} for a description of this sort of quotient map in 
the context of Calabi-Yau spaces defined in weighted projective spaces. 
In particular a weighted $\BP^n_{\bf w}(x_0:\ldots:x_n)$ with the $x_i$ 
satisfying\ 
$(x_0:\ldots:x_n)\ra (\lambda^{w_0}x_0:\ldots:\lambda^{w_n}x_n)$, can 
be described as the quotient 
\bea
\BP^n_{\bf w} = {\BP^n\over \prod_i \BZ_{w_i}} \nonumber
\eea
of the ``isotropic'' $\BP^n(z_0:\ldots:z_n)$ having ${\bf w}=1$, via 
the quotient map\ $x_i=z_i^{w_i}$. In order for the map to be one-to-one, 
\ie\ for an inverse to be well-defined, we must require that the 
coordinates $z_i$ have the identifications $z_i\ra e^{2\pi i/w_i} z_i$.}
\be\label{CFquotient}
{\cal C}^{(flip)} = {{\cal C}\over \prod_i \BZ_{n_i}}
\ee
of the supersymmetric conifold ${\cal C}$ with the action given by 
(\ref{Zn1n2n3n4}), and can be described as the hypersurface 
(\ref{flipconif}) in $\BC^4/\Gamma$ where the discrete group acts 
as (\ref{Zn1n2n3n4}) on the coordinates $z_i$. With the identifications 
(\ref{Zn1n2n3n4}), the relation (\ref{CFquotient}) is one-to-one and the 
inverse is single-valued. Locally the space is isomorphic to the 
supersymmetric conifold. Note that if the $z_i=0$, then we have 
$a,b,c,d=0$. If say $z_1\neq 0$, then, as in Sec.~5.5 of 
\cite{wittenphases}, we 
can take $a\neq 0$ without loss of generality, and then iteratively 
solve (\ref{zin1n2n3n4}) for $c,d,b$, as 
\be
c={z_1^{n_3}\over a^{n_3/n_1}}\ , \qquad d={z_2^{n_4}\over a^{n_4/n_1}}\ , 
\qquad b={z_3^{n_2}\over c^{n_2/n_3}}\ .
\ee
Note that this is a well-defined inverse map, incorporating the 
identifications (\ref{Zn1n2n3n4}) on the $z_i$, as well as the 
relation (\ref{flipconif}). This shows that the quotient relation 
(\ref{CFquotient}) gives a space isomorphic to the toric definition 
using the charge matrix (\ref{Qgen}).

It is useful to note that the conifold relation (\ref{flipconif}) in 
terms of the coordinates $z_i$ subject to orbifolding is defined 
``upstairs'', \ie\ on the covering space. One can realize other 
relations between invariant monomials that are single-valued 
``downstairs'' by using the orbifold action (\ref{Zn1n2n3n4}) as follows. 
Define 
\be
m_1={N\over {\rm lcm} (n_1,n_3)}\ , \ \ 
m_4={N\over {\rm lcm} (n_2,n_4)}\ , \ \ 
m_2={N\over {\rm lcm} (n_1,n_4)}\ , \ \ 
m_3={N\over {\rm lcm} (n_2,n_3)}\ ,
\ee
which are all integers. Then a useful set of invariants defining a 
polynomial ring on this space can be defined as 
\be\label{stuvInv}
s=z_1^{N/m_1} , \ \ \ t=z_4^{N/m_4} , \ \ \ 
u=z_2^{N/m_2} , \ \ \ v=z_3^{N/m_3} , 
\ee
satisfying the ``downstairs'' relation
\be\label{stuvRel}
s^{m_1} t^{m_4} = u^{m_2} v^{m_3}\ ,
\ee
which is essentially the $N$-th power of the conifold equation 
(\ref{flipconif}) ``upstairs'' recast in terms of the invariants 
(\ref{stuvInv}). In general, we also require other monomials 
invariant under the orbifold action (\ref{Zn1n2n3n4}) to describe 
the variety completely ``downstairs''. These monomials can also be 
written as invariants under the original $\BC^*$ action, in terms of 
the underlying variables $a,b,c,d$. The toric variety itself is described 
in terms of the set of relations between these monomials. In general, 
such spaces are not complete intersections of hypersurfaces, \ie\ the 
number of variables minus the number of equations is not equal to 
the dimension of the space. 

Note that the variety\ $s^l t^l = u^k v^k$ \ defined by (\ref{stuvRel}) 
for the case\ $m_1=m_4=l,\ m_2=m_3=k$, obtained by \eg\ $n_1=n_2=k,\ 
n_3=n_4=l$ is on a different footing from similar varieties (and their 
brane duals) studied in \eg\ \cite{gjmukhi9206} \cite{dasguptamukhi9811} 
\cite{uranga9811}. The latter obtain these varieties as supersymmetric 
quotients of the supersymmetric conifold (where the underlying 
variables $a,b,c,d$ have charges $n_i=\pm 1$ with $\sum Q_i=0$). In 
our case on the other hand, the underlying geometry defined by the 
variables $a,b,c,d$ has nontrivial dynamics since it corresponds to 
a charge matrix of the form 
$Q=(\bA{cccc} k & k & -l & -l \eA)$ with\ $\sum Q_i= 2(k-l)\neq 0$. 
The dynamics here is therefore that of an unstable geometry as we 
will see later in Sec.~3.3 using linear sigma models.\\

{\bf Examples}: Consider the flip region $Q=(\bA{cccc} 1 & 2 & -1 & -3 \eA)$ 
of the previous section: here the discrete group $\Gamma=\BZ_2\times \BZ_3$ 
is cyclic, giving $\Gamma=\BZ_6$ and (\ref{Zn1n2n3n4}) effectively 
reduces to\ $( \bA{cccc} z_1 & z_2 & z_3 & z_4 \eA ) \ra 
( \bA{cccc} z_1 & \omega^2 z_2 & \omega^3 z_3 & \omega^5 z_4 \eA ) , 
\ \omega=e^{2\pi i/6}$. 
Then we have $(m_1,m_2,m_3,m_4)=(6,2,3,1)$ and the monomial invariants\ 
$s=z_1, t=z_4^6, u=z_2^3, v=z_3^2, w=z_2 z_4^2, x=z_3 z_4^3$, with the 
relations\ $s^6 t = u^2 v^3,\ w^3=u v,\ x^2=t v,\ st=wx$. 
In terms of the original $\BC^*$ action, these monomials can be written 
as\ $s=ac, t=b^3d^2, u=a^3d, v=bc^2, w=abd, x=b^2cd$, with the 
relations above.\\
For the singularity $Q=(\bA{cccc} 1 & n & -1 & -n \eA)$, the discrete 
group is $\BZ_n\times \BZ_n$. In this case, we obtain\ 
$s=z_1, t=z_4^n, u=z_2^n, v=z_3^n$, satisfying\ $s^n t = uv$. 
Alternatively from the original $\BC^*$ action, we have $s=ac, t=bd, 
u=a^nd, v=bc^n$.

\subsubsection{On deformations of the singularity}

In this section, we want to look for possible 3-cycle 
$\epsilon$-deformations of the form 
\be\label{deformconif}
z_1 z_4 - z_2 z_3 = \epsilon\ ,
\ee
for flip conifolds ($\sum Q_i\neq 0$) with the hypersurface equation
(\ref{flipconif}), with $\epsilon$ chosen a real parameter (by
coordinate changes). Since the flip conifold is a quotient of the
supersymmetric conifold (described as a 3-complex dimensional
hypersurface embedded in the orbifold $\BC^4/\Gamma$ with the conifold
singularity coinciding with the orbifold singularity), one might
imagine that such deformations of the hypersurface singularity are
obstructed by the quotient structure, drawing intuition from the fact
that toric $\BC^d/G, \ d>2$, orbifolds themselves do not admit any
such complex structure deformations \cite{schless}. Following this
logic shows that this is in fact true whenever the singularity is
isolated, \ie\ the $n_i$ are all coprime\footnote{The toric fan of the
singularity encodes information such as whether it is isolated, as we
describe in the next section.}  in which case $\Gamma$ is a cyclic
group. Further analysis using the symmetries of the underlying
variables $a,b,c,d$, or equivalently the structure (\ref{CFquotient})
of these singularities as quotients of the supersymmetric conifold,
shows that in fact 3-cycle deformations are always obstructed, as we
will see below. Another way to state this is: the only complex
structure deformation of the supersymmetric conifold is of the form
(\ref{deformconif}), and the quotient structure of the flip conifold
imposes global obstructions to this possible local
deformation\footnote{This does not however preclude abstract
deformations: see later.}. In what follows, we continue to use the
phrase ``3-cycle'' deformations rather than complex structure
deformations, since we are really only focussing on the local geometry
in the vicinity of the singularity, rather than the full compact
embedding space which is where one would conventionally define
K\"ahler and complex structure deformations.

To begin, let us study flip regions in terms of\ $\nu$, the integers 
$n_i$ in $Q$ being an $ordered$ set. By definition, we have\
$0\leq\nu\leq 4$. $\nu=4$ is only possible when all $n_i=1$ (this is
the supersymmetric conifold), while $\nu=0$ is only possible for all
$n_i\ra\infty$. Thus any finite $n_i$ can at best give $\nu=1, 2, 3$
for integral $\nu$. From the action of the discrete group
(\ref{Zn1n2n3n4}) on the coordinates
(\ref{zin1n2n3n4}) and the structure of the conifold equation
(\ref{flipconif}), we see that whenever $\nu$ is not integral, the
phase $e^{2\pi i\nu}$ given in (\ref{phasenu}) acting on each of the
two terms of the conifold equation is nontrivial, so that the
$\epsilon$-deformation is obstructed. In fact it is straightforward to
show that whenever the singularity is isolated, \ie\ the $n_i$ are
coprime w.r.t. each other, the phase $e^{2\pi i\nu}$ is nontrivial 
and the deformation is obstructed (see the Appendix).

Now if\ $\nu\in\BZ$ as in the select non-isolated cases described 
in the Appendix, then the phase $e^{2\pi i\nu}$ is trivial: this 
would suggest that there is no obstruction to the 3-cycle 
$\epsilon$-deformation of the flip conifold equation 
(\ref{flipconif}). If the phases induced by rotations in $a,b,c,d$, 
were correlated, then the embedding orbifold group $G$ %\equiv\Gamma$ 
would be cyclic, satisfying\ ${\rm det}\ G=e^{2\pi i (2\nu)}=1$, in 
other words, the singularity would be embedded in a supersymmetric 
$\BC^4/G$ orbifold, with the same holonomy ($SU(4)$) as a 
Calabi-Yau 4-fold. For such cases, the analysis of possible 3-cycles 
would coincide mathematically with the analysis of \cite{gopavafa9712} 
who study $S^3/G$ in a supersymmetric context. The 3-cycle in 
these cases is the locus\ $z_4=z_1^*, \ z_3=-z_2^*$, with some 
discrete identifications imposed by the orbifolding, giving $S^3/G$ 
rather than $S^3$, the orbifold $G$ acting freely on the $S^3$. 
For example, consider the conifold singularity with the identifications\ 
$( \bA{cccc} z_1 & z_2 & z_3 & z_4 \eA )\ \ra\ 
( \bA{cccc} -z_1 & -z_2 & -z_3 & -z_4 \eA )$. 
Define new coordinates $w_i$ in terms of linear combinations 
$z_1=w_1+iw_4$ etc, so that the deformed singularity 
(\ref{deformconif}) is now written as\ $\sum_{i=1}^4 w_i^2 = \epsilon$, 
with the 3-cycle being the locus where the $w_i$ are real. The
identifications on the $z_i$ translate to the identifications 
$w_i\ra -w_i$, giving $S^3/\BZ_2$ here. This leads one to ask what 
the corresponding structure of possible deformed singularities is 
(if they exist), in the case when the orbifold group is not cyclic.

With this in mind, recall that the phase rotations of the underlying 
variables $a, b, c, d$, defining the geometric space are independent. 
These are basically the residual, \ie\ unquotiented, torus actions that 
were present in the toric variety defined by the underlying variables 
$a,b,c,d$ (and in the corresponding linear sigma model of the next 
section). Thus consider the path (starting and ending at the same 
point on the space) given by
\be
d \ra e^{2\pi i} d\ , \qquad \qquad a, b, c = {\rm fixed}\ .
\ee
Such paths define isometry directions of the geometry, so the deformed 
hypersurface equation should also respect these if the deformation is 
consistent. However from (\ref{zin1n2n3n4}) (\ref{Zn1n2n3n4}), we see 
that only $z_2, z_4$ acquire a phase and the conifold equation 
(\ref{deformconif}) transforms as 
\be
\epsilon = z_1 z_4 - z_2 z_3\ \ra\  e^{2\pi i/n_4} (z_1 z_4 - z_2 z_3)\ ,
\ee
and similarly for any other such path.
For $n_i\neq 1$, this is consistent only if $\epsilon=0$, \ie\ the 
deformation is obstructed. Another way of seeing that this is sensible 
is to realize that a holomorphic 3-form on the deformed side does not 
exist either, being projected out by the above phase. 

The arguments here have used the ``upstairs'' variables, carefully
implementing the toric symmetries of the underlying geometry described
by the variables $a,b,c,d$: note that these toric symmetries are
equivalent to the structure (\ref{Zn1n2n3n4}) (\ref{CFquotient}) of
these singularities as quotients of the supersymmetric conifold. In
other words, the expression (\ref{deformconif}) is not invariant under
the identifications (\ref{Zn1n2n3n4}) defining the relation
(\ref{CFquotient}), unless the deformation parameter $\epsilon=0$.
These arguments are reasonable for deformations connected to the
singular point since infinitesimal deformations that can be
interpreted ``upstairs'' (\ie\ in the supersymmetric conifold) must be
consistent with the additional toric symmetries (or equivalently the
quotient symmetries) present for flips.  It is noteworthy that these
quotient (or equivalently toric) symmetries thus obstruct the only
complex structure deformation (which is not toric) of the
supersymmetric conifold.

Now we make a few remarks on abstract deformations of these 
singularities which may not allow any interpretation in terms of the 
``upstairs'' structure. It appears that the space of $toric$
deformations (characterized by the fact that the total space including
the deformation still belongs to the toric category) is a large subset
of the so-called versal, or complete, deformation of a singularity 
(see \eg\ \cite{altmann} for deformations of Gorenstein 
singularities\footnote{See also \eg\ \cite{mukhidef9308} for 
semi-universal deformations of ground varieties in the context of $c=1$ 
string theory.}). The versal deformation for a space described as the 
variety $f(x_1,\ldots,x_n)=0$, is given as\ 
$\BC[x_1,\ldots,x_n]/(f,\del f/\del x_1,\ldots,\del f/\del x_n)$, \ie\ 
the polynomial ring modulo the ideals given by the vanishing of $f$ and 
its first partial derivatives $f_i\equiv\del f/\del x_i$: this is 
reasonable to see for infinitesimal deformations since a coordinate 
transformation $x_i'=x+\delta x_i$, of the $x_i$ can absorb first order 
changes as $f(x_i)=f(x_i')-f_i\delta x_i$. It is in general not easy to 
calculate the versal deformation of singularities outside of the formal 
techniques of \eg\ \cite{schless, altmann}. However we can get some 
intuition for this in simple examples. In terms of coordinates 
well-defined ``downstairs'', we can describe \eg\ the simplest flip 
singularity $Q=(\bA{cccc} 1 & 1 & -1 & -2\eA)$ as follows. The 
``upstairs'' coordinates $z_i$ have the $\BZ_2$ identifications 
$( \bA{cccc} z_1 & z_2 & z_3 & z_4 \eA ) \ra^d 
( \bA{cccc} z_1 & -z_2 & z_3 & -z_4 \eA )$. The ``downstairs'' 
coordinates $s=z_1=ac, \ t=z_2^2=a^2d, \ u=z_3=bc, \ v=z_4^2=b^2d, 
\ w=z_2z_4$, generating the monomial ring on the space, satisfy the 
relations\ $f_1=tv-w^2=0, \ f_2=sv-uw=0, \ f_3=tu-sw=0$. This set at 
quadratic order can be regarded as a basis for the ideal of relations 
in this case, since higher order relations follow from these. Note that 
this set of three relations in $\BC^5[s,t,u,v,w]$ is not a complete 
intersection: the toric variety defined by $a,b,c,d$, is 3-complex 
dimensional, which can also be seen by studying the rank of the 
derivative matrix $[{\del f_i\over \del x_j}]$. Alternatively on 
the coordinate patch with nonzero $s,t,u$, we have $w={tu\over s}, 
v={tu^2\over s^2}$ (the third potential relation being an identity), 
so that the space is described by the variables $s,t,u$ (similarly 
for other charts). Thus this singularity is embedded in 
$\BC^5[s,t,u,v,w]$ as the intersection of the supersymmetric 
conifolds $sv-uw=0$ and $sw=tu$, with the $\BC^2/\BZ_2$ orbifold 
$tv-w^2=0$. Let us look for deformations of the two conifolds 
consistent with a deformation of the orbifold: specifically consider\ 
$f_1'=tv-w^2-\epsilon=0,\ f_2'=sv-uw-\lambda=0,\ f_3'=sw-tu-\mu=0$, 
with $\epsilon, \lambda, \mu$ being real. On the $(s,t,u)$ patch, the 
3-cycle in $f_2'$ is the locus $v=s^*, w=-u^*$, while that in $f_3'$ 
is $w=s^*, u=-t^*$, \ie\ we have $s=v^*=w^*=-u=t^*$: this gives 
$\epsilon=(s^*)^2-(s^*)^2=0$ (from $f_1'$), and $\lambda=\mu=2ss^*$, 
which is a $\BP^1$. This abstract deformation (which is not a 3-cycle 
topologically) is difficult to interpret ``upstairs'' in terms of the 
quotient structure: the total space in this case is not toric.
For the general singularity (\ref{Qgen}), writing the explicit
monomial relations and visualizing the geometry is hard, which also
complicates the set of such abstract deformations even along the lines 
of this crude analysis in this simple example. It is also interesting 
to understand the constraints of unbroken worldsheet supersymmetry on 
the quantum descriptions of these 2D theories besides the obstructions 
obtained from the classical geometry analysis here (see Sec.~3.3 and 
the Discussion). The analysis here is of course purely within geometry, 
not accounting for any intrinsically stringy branch \eg\ along some 
flux direction.

To summarize we have looked for 3-cycle deformations using the
quotient structure (\ref{CFquotient}) (\ref{Zn1n2n3n4}): in principle
there could exist abstract deformations of the singularity lying
``outside'' this structure, invisible to the techniques used here. It
would be interesting to understand this better, perhaps using the 
formal techniques of \cite{schless, altmann}.

\subsection{Small resolutions and their dynamics}

Let us now study the dynamics of the small resolutions: these 
necessarily have an inherent directionality whenever $\sum Q_i\neq 0$, 
as we will see from a linear sigma model analysis below.\\
Consider a basis for the \BN\ lattice (and the toric cone in it) given by
\be\label{basisei}
e_2\equiv \phi_2=(1,0,0), \qquad e_3 \equiv \phi_3=(0,1,0), 
\qquad e_4\equiv \phi_4=(0,0,1)\ .
\ee
Then the fourth vertex $\phi_1$ of the cone (see Figure~\ref{figflip}) 
in $\BN$ defining the conifold (see (\ref{fan})) can be rewritten in 
terms of the $e_1,e_2,e_3$ basis as 
\be\label{e1}
e_1\equiv 
\phi_1=-{n_2\over n_1}e_2 + {n_3\over n_1}e_3 + {n_4\over n_1}e_4\ 
\longrightarrow_{n_1=1}\ (-n_2, n_3, n_4)\ .
\ee
(Despite appearances for $n_1\neq 1$, the lattice here is integral as 
we clarify below.) Whenever the $n_i$ are relatively coprime, this 
toric fan does not contain any lattice points in its interior or on 
its ``walls'' --- this is an isolated singularity (see the Example at 
the end of this Section). The four vertices $\{ e_1, e_2, e_3, e_4\}$ 
are coplanar, lying on a hyperplane, if 
\be
0 = {\rm det} (e_1-e_2, e_3-e_2, e_4-e_2) = {n_3+n_4-n_1-n_2\over n_1}\ ,
\ee
in other words, $\sum Q_i = \Delta n = 0$. In this case, the two small
resolutions give identical residual volumes, so that there is no
intrinsic directionality to their dynamics: the fan corresponds to a
spacetime supersymmetric geometry and the two resolutions are related
by a flop, \ie\ a marginal deformation. On the other hand,
non-coplanarity of the $\{ e_i\}$ means that the two small resolutions
(related by a flip transition) give distinct residual subcone volumes
for their corresponding partial blowups, so that one expects an
inherent directionality in the dynamics of the geometry. In both
cases, the two small resolutions are topologically distinct since when
this geometry is embedded in a compact space, the intersection numbers
of various cycles change under the flip/flop.

We can calculate the subcone volumes for each of the small resolutions 
(see Figure~\ref{figflip}):
\bea\label{vol+-}
&& \BP^1_+:\ V_+ = V(0;e_1,e_2,e_3) + V(0;e_1,e_2,e_4) = {n_4\over n_1}+
{n_3\over n_1}\ , \nonumber\\
&& \BP^1_-:\ V_- = V(0;e_2,e_3,e_4) + V(0;e_1,e_3,e_4) = 1 + 
{n_2\over n_1}\ ,
\eea
so that the difference in volumes is
\be\label{DeltaV}
\Delta V = V_- - V_+ = {\Delta n\over n_1}\ .
\ee
This represents the difference in the cumulative degrees of the residual 
singularities for the two small resolutions, and is non-vanishing if 
$\sum_i Q_i = \Delta n \neq 0$. Note that in the normalization where 
the supersymmetric conifold has the charge matrix (\ref{susy1111}), \ie\ 
$Q=(\bA{cccc} 1 & 1 & -1 & -1 \eA)$, we have its residual cumulative 
volumes $V_{\pm}=2$, so that any singularity with $n_1\neq 1$ 
potentially has fractional volumes $V_{\pm}$ --- in the latter cases,
one can choose a different normalization for the supersymmetric
conifold, which then yields an integral $\BN$ lattice. For simplicity,
let us set $n_1=1$ to obtain an integral lattice for arbitrary
$n_2,n_3,n_4$: this gives $\Delta V=\Delta n$. Then the small
resolution decay modes of the flip conifold give rise to the four
residual subcones $C(0;e_i,e_j,e_k)$ which are potentially
$\BC^3/\BZ_M$ singularities.  For all Type II theories, these have
either moduli or tachyons in their spectrum of deformations which can
resolve them completely.  Recalling that $\BC^3/\BZ_2 (1,1,1)$ is a
truly terminal singularity appearing in the spectrum of decay
endpoints in Type 0 unstable orbifolds \cite{drmknmrp}, we obtain a
constraint on when these flip conifolds decay to smooth spaces
assuming the GSO projection is preserved along the flow as it is in
orbifolds \cite{drmknmrp}. For example, $C(0;e_1,e_3,e_4)$ corresponds
to $\BZ_{n_2}(1,-n_3,-n_4)$, so that if $n_2=2$, we must have at least
one of $n_3,n_4$ to be even for this subcone to not be $\BC^3/\BZ_2
(1,1,1)$ \ (for a Type II theory, this condition is automatically met
since the GSO projection \cite{drmknmrp} requires $n_3+n_4=odd$
here). This is a nontrivial requirement: for instance, the singularity
$Q=(\bA{cccc} 1 & 2 & -1 & -3 \eA)$ discussed earlier does not satisfy
this and in fact exhibits the terminal singularity in its decay
endpoints.

In fact we can obtain a general constraint on the $n_i$ from the known 
Type II GSO projection $\sum k_i=even$ \cite{drmknmrp} on the 
$\BC^3/\BZ_M (k_1,k_2,k_3)$ decay endpoints at the IR, if we assume that 
a GSO projection defined in the UV is not broken along the RG flow 
corresponding to the decay. From the Smith normal form algorithm of 
\cite{drmknmrp} (or otherwise), we can see that the various residual 
subcones correspond to the orbifolds\ 
$C(0;e_1,e_2,e_3)\equiv \BZ_{n_4}(1,n_2,-n_3)$, 
$C(0;e_1,e_2,e_4)\equiv \BZ_{n_3}(1,n_2,-n_4)$, and 
$C(0;e_1,e_3,e_4)\equiv \BZ_{n_2}(1,-n_3,-n_4)$, up to shifts of the 
orbifold weights by the respective orbifold orders, since these cannot 
be determined unambiguously by the Smith algorithm\footnote{It is 
useful to note that these $\BC^3/\BZ_M$ orbifolds in the IR are 
isolated if the $n_i$ are all relatively coprime.}. Then we can see 
that each of these orbifolds in the IR of the conifold decay admit a 
consistent Type II GSO projection if 
\be\label{gso}
\Delta n=\sum Q_i = n_1 + n_2 - n_3 - n_4 = even\ .
\ee
If this condition is satisfied, string theory in the flip conifold
spacetime background in question does not have a bulk tachyon at the
top of the tachyon ``hill''. Setting $n_1=1$ for simplicity again,
this implies that $n_2+n_3+n_4=odd$, since $\Delta n$ in (\ref{gso})
is only defined $mod\ 2$. To illustrate this, consider, without loss
of generality, the case $n_2=even$. Then we must have that
$n_3+n_4=odd$, \ie\ one and only one of $n_3, n_4$ is odd. This means
that $C(0;e_1,e_3,e_4)\equiv \BZ_{n_2}(1,-n_3,-n_4)$ automatically
admits a Type II GSO projection. Now say $n_3=odd$. Then the subcone
$C(0;e_1,e_2,e_3)\equiv \BZ_{n_4}(1,n_2,-n_3)$ also admits a Type II
GSO projection, while the subcone
$C(0;e_1,e_2,e_4)\equiv\BZ_{n_3}(1,n_2,-n_4\pm n_3)$, after shifting
one of the weights by the order $n_3$, is also seen to admit a Type II
GSO projection. It is straightforward to show that the other cases are
similarly dealt with. For instance, we can see from this condition
that the singularity $Q=(\bA{cccc} 1 & 2 & -1 & -3 \eA)$ cannot be
Type II, consistent with finding the terminal singularity in its decay
endpoints.

Now consider the dynamics of the small resolutions described by the
$U(1)$ gauged linear sigma model with the four scalars $\Psi\equiv
a,b,c,d$, and a Fayet-Iliopoulos (real) parameter $r$ governing the
vacuum structure: this system has $(2,2)$ worldsheet 
supersymmetry. The fields $\Psi$ transform under $U(1)$ gauge 
transformations with the charge matrix $Q_i$ as
\be\label{U1gt}
\Psi_i \ra e^{i Q_i\beta} \Psi_i, \qquad\qquad Q_i = (n_1, n_2, -n_3, -n_4)\ ,
\ee
$\beta$ being the gauge parameter. As in the description of orbifold 
flips in Sec.~2, the action for the GLSM is 
\be
S = \int d^2 z\ \biggl[ d^4 \theta\ \biggl( {\bar \Psi_i} e^{2Q_i V} 
\Psi_i - {1\over 4e^2} {\bar \Sigma} \Sigma \biggr) + \Rea\biggl( i 
t\int d^2 {\tilde \theta}\ \Sigma  \biggr) \biggr]\ ,
\ee
where $t = ir + {\theta\over 2\pi}$ , $\theta$ being the $\theta$-angle 
in $1+1$-dimensions. The classical vacuum structure can be found, as 
in Sec.~2, by the bosonic potential (\ref{bospot}): as in that case, 
the $\sigma$ fields have zero vevs for nonzero $r$, and the classical 
vacuum structure is described by the D-term equation
\be
-{D\over e^2} = n_1|a|^2 + n_2|b|^2 - n_3|c|^2 - n_4|d|^2 - r = 0\ ,
\ee
divided by $U(1)$, from which one can realize the two small resolutions 
(rank-2 bundles over $\BP^1_{\pm}$) as manifested by the moduli space 
for the single FI parameter ranges $r\gg 0$ and $r\ll 0$. The geometry of 
this space can be understood thinking of this as a symplectic quotient. 
Firstly consider the phase $r\gg 0$: then one of $a, b$ is nonzero. 
Consider for simplicity the case where $n_i$ are all coprime. Then 
$z_+={a^{n_2}\over b^{n_1}}$, invariant under the $U(1)$ defines a 
coordinate on the $\BP^1$ on the patch where 
$b\neq 0$ (with $z_-={1\over z_+}$ the corresponding coordinate on the 
patch where $a\neq 0$). The full space, a rank-2 bundle over $\BP^1$, 
can then be described in terms of the invariants\ 
$p_-=a^{n_3}c^{n_1}, \ p_+=b^{n_3}c^{n_2}, \ q_-=a^{n_4}d^{n_1}, \ 
q_+=b^{n_4}d^{n_2}$, satisfying the relations\ 
$p_-^{n_2}=p_+^{n_1} z_+^{n_3}, \ q_-^{n_2}=q_+^{n_1} z_+^{n_4}$, 
which describes the rank-2 bundle locally as a variety embedded in 
$\BC^2(p_-,p_+)\times\BC^2(q_-,q_+)$. 
This is the small resolution for the phase $r\gg 0$. The residual 
$\BC^3/\BZ_M$ singularities in this phase are realized by looking at 
the regions in moduli space where only one of $a, b$, acquires a 
vacuum expectation value: for instance, a vev for $b$ alone Higgses 
the $U(1)$ down to $\BZ_{n_2}$ with the chart $(a,c,d)$, \ie\ the 
subcone $C(0;e_1,e_3,e_4)$. \\
The other small resolution for the phase $r\ll 0$ can be similarly 
studied in terms of the invariant coordinates 
$z_+'={c^{n_4}\over d^{n_3}}\equiv \BP^1$, and 
$p_-'=c^{n_1}a^{n_3}, \ p_+'=d^{n_1}a^{n_4}, \ \ q_-'=c^{n_2}b^{n_3}, \ 
q_+=d^{n_2}b^{n_4}$, satisfying the relations\ 
${p_-'}^{n_4}={p_+'}^{n_3} {z_+'}^{n_1}, \ 
{q_-'}^{n_4}={q_+'}^{n_3} {z_+'}^{n_2}$, which describes the bundle 
locally as a variety in $\BC^2(p_-',p_+')\times\BC^2(q_-',q_+')$. 

This describes how the resolved flip conifold phases, rank-2 bundles 
over $\BP^1_{\pm}$, are embedded in general locally as 3-complex 
dimensional spaces in $\BP^1_{\pm}\times\BC^2\times\BC^2$.
Note that for special cases, this simplifies to the structure of an 
${\cal O}(-l_1)\oplus {\cal O}(-l_2)$ bundle over $\BP^1$, for 
appropriate integers $l_1,l_2$, as we will see in an example later. 
Perhaps one can also usefully describe the total space in general as 
a bundle analog of an orbifold since the fibre coordinates 
$p_{\pm}, q_{\pm}$ (as well as $p_{\pm}', q_{\pm}'$) above have 
identifications. For the case where the $n_i$ are not all coprime, 
one defines coordinates on the $\BP^1$ after eliminating the g.c.d. 
and so on, and similarly for the fibre coordinates. 

Analogous to (\ref{rflowOrbs}) for orbifolds, the parameter $r$ has a 
1-loop renormalization given by
\be\label{rflow}
r = \biggl( \sum_i Q_i\biggr) \log {\mu\over \Lambda}
= ( \Delta n ) \log {\mu\over \Lambda}
= (\Delta V) \log {\mu\over \Lambda}\ ,
\ee
(setting $n_1=1$ here for simplicity) identifying the coefficient of the 
logarithm with the volume difference (\ref{DeltaV}). For $\sum_iQ_i=0$, 
there is no 1-loop renormalization and $r$ is expected to correspond 
to a marginal operator in the corresponding string conformal field 
theory. Note that the two phases are still topologically distinct 
geometries in general, as described above. For $\sum_iQ_i\neq 0$, 
the GLSM RG flow (\ref{rflow}) drives the system towards the phase 
corresponding to smaller $\BN$ lattice volume, the residual volumes 
for the two resolutions and their differences being given in 
(\ref{vol+-}) (\ref{DeltaV}), much like the corresponding flow 
(\ref{voldiffOrbs}) (\ref{rflowOrbs}) in orbifold flips described in 
Sec.~2. This shows that the conifold dynamically evolves towards the 
less singular, and therefore more stable, small
resolution, similar to flips arising in unstable orbifolds
\cite{drmkn}. In particular, if one sets up initial conditions for the
geometry to lie in the less stable small resolution, then small 
fluctuations will force the system to evolve through a flip transition:
the mild topology change here, with the blown-down 2-cycle $\BP^1_+$ 
and the blown-up 2-cycle $\BP^1_-$ changing the intersection 
numbers\footnote{The changes in \eg\ the triple intersection numbers 
of divisors can be seen directly from the toric fan in 
Figure~\ref{figflip} using standard toric geometry calculations.} of 
various cycles, is dynamically mediated by closed string tachyon 
instabilities in the geometry. Our discussion here is restricted to 
the large $r$ semiclassical regions of the geometry where quantum
corrections are small, but this is sufficient insofar as an
understanding of the long (RG) timescale dynamics of the system is
concerned. It is important to note that we have decoupled gravity 
here, as for orbifold flips in Sec.~2: this assumption on the 
validity of the GLSM for long timescales is equivalent to the 
reasonable assumption that operator mixing along RG trajectories is 
negligible insofar as an understanding of the phase structure at the 
onset of localized tachyon condensation is concerned (the GSO 
projection being preserved is a consistency check). From a spacetime
point of view, operator mixing is expected on sufficiently long
timescales where one must pass to an appropriate gravity description.

We can calculate the bosonic potential (\ref{twistedWpot}) from the 
quantum twisted chiral superpotential (\ref{twistedW}) as
\be\label{bospotF}
U(\sigma) = {e^2\over 2} \bigg| i{\hat \tau} - 
{\sum_i Q_i \over 2\pi} (\log (\sqrt{2} Q_i \sigma/\Lambda) + 1) \bigg|^2\ .
\ee
As for orbifold flips \cite{drmkn}, the potential has a $|\log\sigma|^2$ 
growth along the relevant direction which is the single direction here: 
thus the quantum corrections serve to smoothing out the transition
avoiding the classical singularity at $r=0$. It would appear that
there exist isolated Coulomb branch $\sigma$-vacua in the infrared of
this system too, as in unstable orbifolds: perhaps the analysis there
\cite{martinecmoore, moore0403, immrp0501, moore0507, immrp0507} can
also be used to give insight here\footnote{Note also the analysis
\cite{egovindtj0504} of fractional branes in a nonsupersymmetric
$\BC^2/\BZ_M$ orbifold via an embedding thereof in a higher
dimensional supersymmetric singularity $\BC^3/\Gamma$. See also
\cite{sarkar0407}.}.

Note that we have not actually constructed the tachyonic states
explicitly in a conformal field theory in this discussion: we have
resorted to indirect means such as the linear sigma model\footnote{For
flip regions arising in nonsupersymmetric $\BC^3/\BZ_N$ orbifolds, one
can construct the tachyons explicitly as twisted sector excitations in
the orbifold conformal field theory as in \cite{drmknmrp}
\cite{drmkn}.}. It would be interesting to construct \eg\ Gepner-like
models describing nonsupersymmetric orbifolds of Calabi-Yau spaces
which develop flip conifold singularities, with a view to more
concretely realizing closed string tachyons in such geometries.

Finally, our assumption that the GSO projection is not broken along
the RG flow corresponding to the decay is not unreasonable physically.
For instance consider the case when the GSO is broken, \ie\ the flip
conifold background $\BR^{3,1}\times {\cal C}^{(flip)}$ has a bulk
tachyon while the $\BC^3/\BZ_M$ orbifold endpoints in the IR do
not. Then along the localized tachyon RG flow described by the GLSM,
the bulk tachyon must somehow disappear. This seems unlikely since in
spacetime, the decay of this conifold is via a small resolution
involving a $\BP^1$ expanding to large size with potential orbifold
singularities whose spatial separation from each other grows at the
rate the volume of the 2-sphere grows. It would seem inconsistent to
have a delocalized tachyon suddenly disappear (or appear) in the
process of condensation of localized tachyons (which is rendered
further credence by the fact that $(2,2)$ worldsheet supersymmetry is
not broken in this GLSM): on the other hand, a more conservative view
is that the GLSM is perhaps not a good tool to describe tachyon 
condensation phenomena where the GSO projection is broken. In any
case, it would be interesting to recover the GSO constraint
(\ref{gso}) more directly, \eg\ modular invariance of the partition
function of an appropriate Gepner-like model. \\

{\bf Example}: Consider the singularity\ 
$Q=(\bA{cccc} 1 & 1 & -2 & -2\eA)$. The coordinates $z_i$ have the 
$\BZ_2\times \BZ_2$ identifications 
$( \bA{cccc} z_1 & z_2 & z_3 & z_4 \eA ) \ra^c 
( \bA{cccc} -z_1 & z_2 & -z_3 & z_4 \eA )$ and 
$( \bA{cccc} z_1 & z_2 & z_3 & z_4 \eA ) \ra^d 
( \bA{cccc} z_1 & -z_2 & z_3 & -z_4 \eA )$. The ``downstairs'' coordinates 
$x_1=z_1^2=a^2c, \ x_2=z_2^2=a^2d, \ x_3=z_3^2=b^2c, \ x_4=z_4^2=b^2d, 
\ x_5=z_1z_3=abc, \ x_6=z_2z_4=abd$, generating the monomial ring
describe the singularity as the intersection of the four relations\ \ 
$x_1x_4=x_2x_3, \ x_5x_6=x_1x_4, \ x_5^2=x_1x_3, \ x_6^2=x_2x_4$,\ in 
$\BC^6[x_1,\ldots,x_6]$.\\
This singularity admits a Type II GSO projection ($\sum Q_i=even$). 
From (\ref{basisei}) (\ref{e1}), we see that the cone
(Figure~\ref{figflip}) is defined by the lattice points $e_2,e_3,e_4$
and $e_1=(-1,2,2)$. Then $V_-=2, V_+=4$, and the less singular
resolution is $\BP^1_-$. A signature that this is a non-isolated
singularity is the interior lattice point $(0,1,1)={e_1+e_2\over 2}$
lying on the wall $\{0,e_1,e_2\}$ defining the $\BP^1_+$ resolution:
note that this point lies along $\{0,e_3,e_4\}$ defining the less
singular $\BP^1_-$ resolution but ``above'' the cone
$\{0;e_1,e_2,e_3,e_4\}$. The residual subcones under the resolution
$\BP^1_+$ are $C(0;e_1,e_2,e_4)\equiv \BZ_2 (1,1)$,
$C(0;e_1,e_2,e_3)\equiv \BZ_2 (1,1)$\ (which are both supersymmetric,
the point $(0,1,1)$ above defining the marginal blowup mode), with the
other two subcones, under $\BP^1_-$, being smooth ($volume=1$).  We
have $\sum Q_i\neq 0$ here and the GLSM dictates an RG flow to the
$\BP^1_-$ resolution. This corresponds to the total space of an ${\cal
O}(-2)\oplus {\cal O}(-2)$ bundle over $\BP^1_-$, realized by defining
$z_+={a\over b}, \ p_-=a^2c, p_+=b^2c, q_-=a^2d, q_+=b^2d$, satisfying
$p_-=p_+ z_+^2, \ q_-=q_+ z_+^2$. The other resolution is an ${\cal
O}(-1)\oplus {\cal O}(-1)$ bundle over $\BP^1_+$, realized by
$z_+'={c\over d}, \ p_-'=ca^2, p_+'=da^2, q_-'=cb^2, q_+'=db^2$,
satisfying $p_-'=p_+' z_+', \ q_-'=q_+' z_+'$.

\section{Discussion}

We have studied the local geometry and dynamics of flip conifold
singularities in this work. Along the lines of \cite{klebwitt,
morrisonplesserHorizons} for the supersymmetric conifold, we can
attempt a construction of the quiver theory \cite{douglasmoore} on a
D3-brane probe near a flip conifold singularity. Consider a
$U(1)\times U(1)$ gauge theory and bifundamental scalars $A_1, A_2,
B_1, B_2$ carrying charges $(n_i,-n_i),\ i=1\ldots 4$. These are
neutral under the diagonal $U(1)$, the free photon on the
D3-brane. The moduli space of vacua is determined by the D-term
condition
\be
D = n_1|A_1|^2 + n_2|A_2|^2 - n_3|B_1|^2 - n_4|B_2|^2 - \rho = 0\ ,
\ee
divided by $U(1)$, $\rho$ being the Fayet-Iliopoulos coupling on 
the D-brane worldvolume. Since $\sum Q_i\neq 0$, it is not possible 
to write a superpotential of the form 
$\epsilon^{ij}\epsilon^{kl} {\rm Tr} (A_i B_k A_j B_l)$ that is
invariant under the $U(1)$ action: this is reminiscent of a similar
statement for a superpotential like ${\rm Tr} X[Y,Z]$ for
nonsupersymmetric orbifolds, and is not surprising since spacetime
supersymmetry is broken: there would of course be a bosonic potential
energy which would be interesting to nail down. We imagine that a
stack of $k$ D3-branes sitting at the singularity (either orbifold or
conifold) would admit an $AdS_5\times X^5$ supergravity description,
where $X^5$ is some appropriate compact manifold. For isolated
singularities, there is no place where light localized tachyonic
states can emerge so that one expects a tachyon-free large $k$ (\ie\
large flux) description. This suggests that the dual gauge theory is
insensitive to the closed string tachyonic instabilities, and thus
stable, at leading order in ${1\over k}$ . It is tempting to speculate
the existence of nontrivial gauge theories by appropriate tuning of 
the parameters in the geometry and the number $k$ of D3-branes (as 
well as possible fractional branes), as nonsupersymmetric duals to 
the supergravity geometries. Subleading ${1\over k}$ effects might 
then encode the tachyonic instabilities in the gauge theory. It would 
be interesting to develop this further and understand possible 
tachyonic open-closed dualities \cite{WiP}.
\begin{figure}
\bc
\epsfig{file=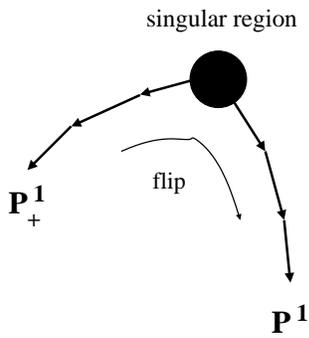, width=4cm}
\caption{A heuristic picture of the possible directions for the evolution 
of a flip conifold singularity, the two small resolutions $\BP^1_{\pm}$: 
possible 3-cycle deformations are obstructed. The direction of the flip 
shown here is for the case when $\BP^1_-$ is the stable small resolution.}
\label{figconif}
\ec
\end{figure}

In general, one imagines such a geometry to be embedded as a local
singularity in some compact space: for instance one expects that
appropriate nonsupersymmetric orbifolds of Calabi-Yau spaces can be
embedding spaces (although perhaps only for the low-lying
singularities, \ie\ small $n_i$). We expect that the local analysis
here is largely independent of the details of the global embedding. In
this case, the phenomena discussed above lead to dynamical topology
change of the geometry, with the intersection numbers of cycles
changing, as for orbifold flips \cite{drmknmrp, drmkn}. Our analysis 
of the physics here has been somewhat indirect, using linear sigma
models: in particular it would be useful to construct the localized
tachyons explicitly \cite{WiP}. From the above analysis, we make the
following statements and speculations (see Figure~\ref{figconif} for a
heuristic picture), with intuition based on spacetime perspectives of
the physics rather than \eg\ worldsheet RG flows.

Given the obstructions to 3-cycle deformations, the phases of flip 
conifolds simply consist of the small resolutions, bundles over 
$\BP^1_{\pm}$ with possible residual $\BC^3/G$ singularities. Such a 
singularity decays to its stable small resolution: since the dynamics
forces the 2-cycles to expand in time (as is seen from the flow to
large K\"ahler parameter $r$ in the GLSM of the previous section),
this means that these unstable conifolds basically decay to smooth
spaces, since the possible residual orbifold singularities if unstable
will themselves decay to smooth spaces too \cite{drmknmrp}. There is
an inherent directionality in time in the dynamics here. Say at early
times one sets up the system in the unstable small resolution (which
is a semiclassical phase if the 2-cycle volume is large, \ie\ large 
$r$): then the dynamics dictates a flip transition towards the stable 
one executing mild topology change\footnote{We are using ``mild'' and
``strong'' along the lines of \eg\ \cite{greeneCY}.} in the process.
There appears to be no analog here of the phenomena associated with
strong topology change \cite{strominger} \cite{greemorrstro}: the
small resolutions simply replace one 2-sphere with another so that
while the intersection numbers of various cycles in a compact
embedding in general change under the corresponding flip transition,
the Euler number does not. We have been largely using classical
geometry to study deformations so far: one could therefore ask if
conifold transitions could occur nonperturbatively since spacetime
supersymmetry is broken. In other words, we ask whether new 3-cycle
branches open up nonperturbatively close to the transition point
causing drastic tears in spacetime. While we cannot rule it out, this
seems unlikely since $(2,2)$ worldsheet supersymmetry is unbroken and
worldsheet instantons ensure that there is a $|\log\sigma|^2$ growth
(\ref{bospotF}) in the potential energy given by the quantum twisted
chiral superpotential.  Thus unlike the supersymmetric conifold, there
are no singularities in the Higgs branch where new 3-cycle (continuous
Coulomb) branches can open up along the RG flow direction (which also
suggests that the abstract deformations described at the end of
Sec.~3.2.1 are physically unlikely): in other words, the quantum
2-cycle volumes are not vanishing and one does not expect sufficiently
light wrapped brane states.  It is worthwhile noting that the dynamics
of a flip is quite different from a time-varying flop mediated by say
a slowly varying modulus: by the (RG) time the evolution of the
geometry crosses over between phases, the situation is far from a slow
variation and is rather a rapid transition between phases of distinct
topology.  Thus the rate at which the geometry is evolving in time (in
spacetime) ``near'' the transition itself is large, and the region
near the singularity where quantum corrections are large is a
transient intermediate state. It would be interesting to investigate
in detail the dynamics here in the context of a compact embedding with
fluxes, in part to understand if there are analogs here of ``moduli''
trapping \cite{evaModTrap0403} and other phenomena.

Consider now the dynamics of the compact space in which unstable
singularities are embedded: a nonsupersymmetric orbifold of $\BC\BP^3$
would typically contain isolated $\BC^3/\BZ_N$ singularities. From a
linear sigma model analysis, we see that $\BC\BP^3$ has a tendency to
spontaneously shrink its overall volume (a nice discussion of this can
be found in \eg\ \cite{wittenIAS}). Then one imagines that the
evolution\footnote{Note that $\BC\BP^3$, with nonzero curvature, is
not a solution by itself to the string equations of motion: what we
imagine here is that an appropriate solution can be constructed in
terms of a nontrivial FRW-like cosmology in the remaining 4D
spacetime, where the evolution of the 4D scale factor reflects that of
the internal space, somewhat along the lines of the cosmological 
solutions in \eg\ \cite{cdnsz0501}.}  of an orbifold of $\BC\BP^3$ 
would depend on the competition between this tendency to spontaneously
shrink and the effects of possible tachyonic cycles expanding as a
result of condensation of closed string tachyons localized to the
isolated $\BC^3/\BZ_N$ singularities: a naive expectation is that the
overall tendency to collapse perhaps dominates the expanding tachyons
essentially since the collapse is a bulk effect whereas the tachyons
are localized objects associated to cycles of some nonzero 
codimension, resulting in a collapsed $\BC\BP^3$ as the late time
endpoint\footnote{This argument is perhaps too glib when the overall
size of the $\BC\BP^3$ shrinks to substringy length scales, where a
more careful analysis going beyond geometry is required.}. However now
imagine turning on some fluxes: then the effective mass for a twisted
state tachyon $T$ increases as\ $m_{eff}^2=-m_0^2+|F|^2$, from the
coupling of the fluxes to the twisted sector states \cite{eva0407}.
Thus maybe more interesting and useful endpoints could result if
appropriate fluxes stabilize both localized tachyon masses and the
overall dynamical volume.  A Calabi-Yau orbifold embedding of a
nonsupersymmetric $\BC^3/\BZ_N$ singularity has no obvious such
tendency to spontaneously shrink so one might naively expect that it
simply evolves spontaneously towards a large volume limit with the
expanding tachyonic cycles. On the other hand, one might expect that
an unstable flip conifold can be embedded in an appropriate
nonsupersymmetric orbifold of a Calabi-Yau that develops a
supersymmetric conifold singularity, with the orbifold action on the
local supersymmetric conifold singularity resulting locally in the
flip conifold singularity in question. However there may be
constraints on the $n_i$ defining the singularity for the existence of
such an embedding in a space that is locally Calabi-Yau (but with
global identifications): \ie\ not all such local singularities may
admit locally supersymmetric string compactifications. It would be
worth exploring the dynamics of these systems further.

\vspace{15mm}

{\small {\bf Acknowledgments:} It is a pleasure to thank D.~Morrison
for collaboration in the incipient stages of this work as well as
several useful discussions. I have also benefitted from useful
discussions with S.~Ashok, C.~Beasley, A.~Dabholkar, R.~Gopakumar,
S.~Govindarajan, T.~Jayaraman, S.~Mukhi, N.~Nitsure, A.~Sen and
C.~Vafa. I'd like to thank the Organizers of the ``Mathematical
Structures in String Theory'' Workshop and the hospitality of the KITP
Santa Barbara, USA, and the Harvard Theory Group, USA, as this paper
was being finalized. This research was supported in part by the
National Science Foundation under Grant No. PHY99-07949.}

\vspace{10mm}

\appendix
\section{More on deformations}

Consider a nonminimal flip region, \ie\ at most one of the $n_i$ is 
equal to one: say $n_1=1$. We now look for cases when\ 
$\nu = 1+{1\ov n_2}+{1\ov n_3}+{1\ov n_4}\in\BZ$, \ie\ the phase is 
trivial. Since $\nu>1$, we need to look for $n_i$ such that $\nu=2,3$.
We see that any two integers $n_3,n_4$ satisfying\ 
$n_2={n_3n_4\over (\nu-1)n_3n_4-n_3-n_4}$ \ such that $n_2\in\BZ^+$ 
give $\nu=2,3$ and therefore give such nonminimal flip regions.
These have charge matrices (with $\nu=2,3$ a parameter) 
\be\label{QwithS3}
Q=(\bA{cccc} 1 & {n_3 n_4\ov (\nu-1)n_3n_4-n_3-n_4} & -n_3 & -n_4 \eA) \ 
\equiv\ (\bA{cccc} 1 & K & -n_3 & -n_4 \eA)\ , \qquad n_3,n_4\in\BZ^+\ .
\ee
For this singularity to be isolated, we must have $n_3, n_4, K$ to 
be positive integers coprime w.r.t. each other. Let $n_3, n_4$ to be 
written via their prime factors as 
$n_3 = p_1.p_2...p_k, \  n_4 = q_1.q_2...q_l$ . For $n_3, n_4$ to be 
coprime, all the $p_i, q_j$ must be coprime. Then we want
\be
K = {n_3 n_4\ov (\nu-1)n_3n_4-n_3-n_4} =
{(p_1 p_2 \ldots p_k)\ (q_1 q_2\ldots q_l)\ov 
((\nu-1)p_1\ldots p_k \cdot q_1\ldots q_l)
- (p_1\ldots p_k) - (q_1\ldots q_l)}
\ee
to be a positive integer and coprime to the two factors in the numerator.
If the denominator divides the numerator, then it must be that some of the 
numerator $p_i,q_j$ factors cancel (using prime factorization of the 
denominator). Then we have $K=\prod (p_i' q_j')$, the residual product 
(deleting the cancelled primes). Thus $K$ has common factors with 
$n_3, n_4$, \ie\ the residual primes $p_i', q_j'$ --- hence non-isolated.
More generally consider none of the $n_i$ to be equal to one. Then for 
say $\nu=1$, using (\ref{phasenu}) we have\ 
$n_4={n_1n_2n_3\over n_1n_2n_3-n_2n_3-n_3n_1-n_1n_2}$ , so that if 
$n_4\in\BZ$, then as above it is not coprime with $n_1, n_2, n_3$, 
hence not isolated. Similarly, if $\nu=2,3$, we have\ 
$n_4=(n_1 n_2 n_3)/(\ldots)$, hence not isolated. 
Similarly consider the case when at least two of the $n_i$ are equal to 
one, \ie\ a minimal flip region\footnote{For instance, in the 
singularity of Sec.~2 and Sec.~3.1, we had 
$Q = (\bA{cccc} 1 & 2 & -1 & -3 \eA)$.}, say $n_1,n_2=1$, with the 
charge matrix 
\be\label{11n3n4}
Q = \left( \bA{cccc} 1 & 1 & -n_3 & -n_4 \eA \right)\ .
\ee
For the phase to be trivial, we want\ 
$\nu=2+{1\ov n_3}+{1\ov n_4}\in\BZ$. The pair\ $(n_3,n_4)=(2,2)$ \ 
giving\ $\nu=1$ is the charge matrix\ $Q=(\bA{cccc} 1 & 1 & -2 & -2 \eA)$, 
with $\sum Q_i=-2\neq 0$. More generally, we can see that this is the 
unique such region in this case: since $\nu>2$ here, $\nu=3$ implies 
${1\ov n_3}+{1\ov n_4}=1$,\ \ie\ $n_4={n_3\over n_3-1}$, which is 
integral only if $(n_3-1)$ divides $n_3$, \ie\ $n_3=2,n_4=2$. This 
shows that there are no deformations whenever the $n_i$ are not of 
this form.

\end{document}